\documentclass[british,english,aps,twocolumn, superscriptaddress, prl]{revtex4-1}
\usepackage[LGR,T1]{fontenc}
\usepackage[latin9]{inputenc}
\usepackage{array}
\usepackage{textcomp}
\usepackage{graphicx}
\usepackage{subscript}

\makeatletter

\DeclareRobustCommand{\greektext}{%
  \fontencoding{LGR}\selectfont\def\encodingdefault{LGR}}
\DeclareRobustCommand{\textgreek}[1]{\leavevmode{\greektext #1}}
\DeclareFontEncoding{LGR}{}{}
\DeclareTextSymbol{\~}{LGR}{126}
\providecommand{\tabularnewline}{\\}
\newcommand{\lyxdot}{.}

\makeatother

\usepackage{babel}
\begin{document}
\title{Measuring the Proton Selectivity of Graphene Membranes}

\author{Michael I. Walker}
\affiliation{Cavendish Laboratory, University of Cambridge, J.J. Thomson Avenue, Cambridge CB3 0HE, United Kingdom}
\author{Philipp Braeuninger-Weimar}
\affiliation{Department of Engineering, University of Cambridge, Cambridge CB3 0FA, United Kingdom}
\author{Robert S. Weatherup}
\affiliation{Department of Engineering, University of Cambridge, Cambridge CB3 0FA, United Kingdom}
\author{Stephan Hofmann}
\affiliation{Department of Engineering, University of Cambridge, Cambridge CB3 0FA, United Kingdom}
\author{Ulrich F. Keyser}
\affiliation{Cavendish Laboratory, University of Cambridge, J.J. Thomson Avenue, Cambridge CB3 0HE, United Kingdom}
\email{ufk20@cam.ac.uk}

\begin{abstract}
By systematically studying the proton selectivity of free-standing graphene membranes in aqueous solutions we demonstrate that protons are transported by passing through defects. We study the current-voltage characteristics of single-layer graphene grown by chemical vapour deposition (CVD) when a concentration gradient of HCl exists across it. Our measurements can unambiguously determine that H\textsuperscript{+} ions are responsible for the selective part of the ionic current. By comparing the observed reversal potentials with positive and negative controls we demonstrate that the as-grown graphene is only weakly selective for protons. We use atomic layer deposition to block most of the defects in our CVD graphene. Our results show that a reduction in defect size decreases the ionic current but increases proton selectivity. 
\end{abstract}

\maketitle

The selective transport of ions across graphene membranes in aqueous
solutions is a key issue for high profile applications such as water
desalination and proton exchange membranes. A number of studies have
reported simulations~\cite{Sint2008,Kang2014}, measurements of filtration
through graphene membranes~\cite{Surwade2015,O'Hern2012,O'Hern2014,Koenig2012}
and the translocations of macromolecules~\cite{Garaj2010a,Schneider2010,Merchant2010}.
Proton transport across graphene membranes is especially interesting
as recent reports have suggested proton selective transport across
intact graphene membranes~\cite{Hu2014}. However, there is little
consensus regarding the mechanisms of transport and the appropriate
experiments to conclude that a graphene membrane is proton selective.

Determining which ions cross a membrane is a problem that has been
extensively studied in the field of ion channels. Selective transport
of ions and macromolecules across biological membranes is critical
to the operation of biological systems. Since the work of Hodgkin
and Huxley in 1939 these have traditionally been investigated by electrophysiology
techniques and single channel measurements~\cite{Hodgkin1939}. These
methods can be used to determine selective transport in the presence
of leakage and multiple transport pathways.

\begin{figure}
\includegraphics{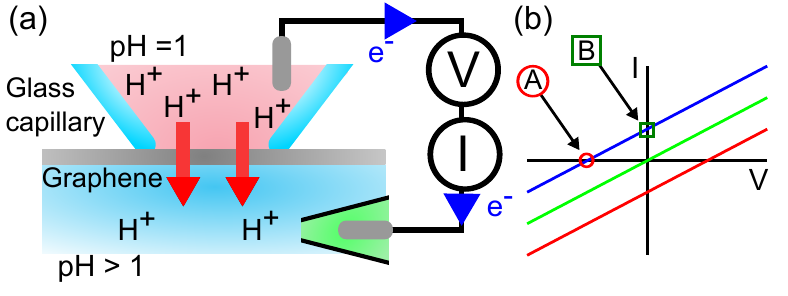}

\protect\caption{(a) A schematic of our experiment showing a glass nanocapillary sealed
with a graphene membrane separating HCl solutions of different concentrations.
Ag/AgCl electrodes in the reservoir and the capillary connect the
experiment to an amplifier which applies a voltage and measures the
current. The arrows show the diffusion current of protons due to the
concentration gradient. (b) An illustration of the expected changes
in the I-V characteristic when a concentration gradient across a selective
membrane drives a diffusion current at zero potential (B) and the
reversal potential required to stop the current (A). \label{fig:Fig1}}
\end{figure}

An established and immediate measurement of ion selectivity is to
set up a concentration gradient over the membrane as shown in Fig~\ref{fig:Fig1}(a).
This creates a driving force for diffusion for both the positive and
negative ions. However if one of the ions can pass through the membrane
more easily than the other then there will be a measurable net current
flow across the membrane when the potential is zero, indicated by
point B in the idealised current (I)~-~voltage (V) characteristic
in Fig~\ref{fig:Fig1}(b)~\cite{Hille2001}. This current can be
stopped by applying an opposing electric field. The corresponding
voltage is called the reversal potential and is illustrated as point
A in Fig~\ref{fig:Fig1}(b). The voltage required is predicted by
the Nernst equation, but deviations from the expected value can be
used to assess the importance of leakage currents~\cite{Hille2001}.

Here we present experiments to study the selectivity of graphene membranes
to protons in aqueous solutions. By systematically measuring the competition
between H\textsuperscript{+} and Cl\textsuperscript{-} ions to cross
single layer graphene membranes we can distinguish which ion is carrying
the current. This is critical because all measurements across a membrane
will measure a leakage current. This current could be passing via
the seal around the membrane, directly though the substrate or, importantly
for graphene, be due to defects in the sample. To dissociate the selective
effect from a leakage current we create a concentration difference
and investigate the I-V characteristics. From the reversal potential
we can determine the extent to which our CVD graphene membranes are
selective to protons. To establish the effect of defects we block
the pores using atomic layer deposition (ALD) and as a positive control
present results using the proton selective membrane Nafion.

We seal graphene membranes on to the tips of pulled glass nanocapillaries
with diameters of 180~nm and 2~\textgreek{m}m using the method developed
in~\cite{Walker2015} and illustrated in Fig~\ref{fig:Fig1}(a).
The graphene adheres to the tips of the capillaries so that the solution
in the reservoir can be exchanged for a different concentration. Ag/AgCl
electrodes in the capillary and reservoir carry the current which
is measured using an Axopatch 200B amplifier \emph{(Molecular Devices)}
used to take I-V curve measurements. Graphene was grown by chemical
vapour deposition in an Aixtron BM Pro (4~inch) reactor, using 25~\textgreek{m}m
thick Cu foil (Alfa Aesar, 99.8\%) as the catalyst and CH\textsubscript{4}
(diluted in Ar and H\textsubscript{2}) as the precursor at 1050\textdegree C~\cite{Hofmann2015}.
Our single-layer graphene significantly impedes the flow of current~\cite{Walker2015}.
We keep the solution inside the capillaries at 0.1M HCl and vary the
concentration of the HCl solution in the reservoir. It is important
to ensure the electrode potentials remain constant despite changing
the concentration of the reservoir. To achieve this we use agarose
coated electrodes made up in 0.1M KCl solution (Fig~\ref{fig:Fig1}(a)).
Our negative control experiments, using bare capillaries, show that
the solution can be exchanged without a significant current being
induced when no membrane is present. As a positive control we will
present results using the proton selective membrane Nafion~\cite{Mauritz2004};
a commercial proton exchange membrane. It is highly conductive to
protons (cations) but blocks anions. The Nafion is 100~\textmu m
thick and is contacted in the same way as the graphene.

\begin{figure}
\includegraphics{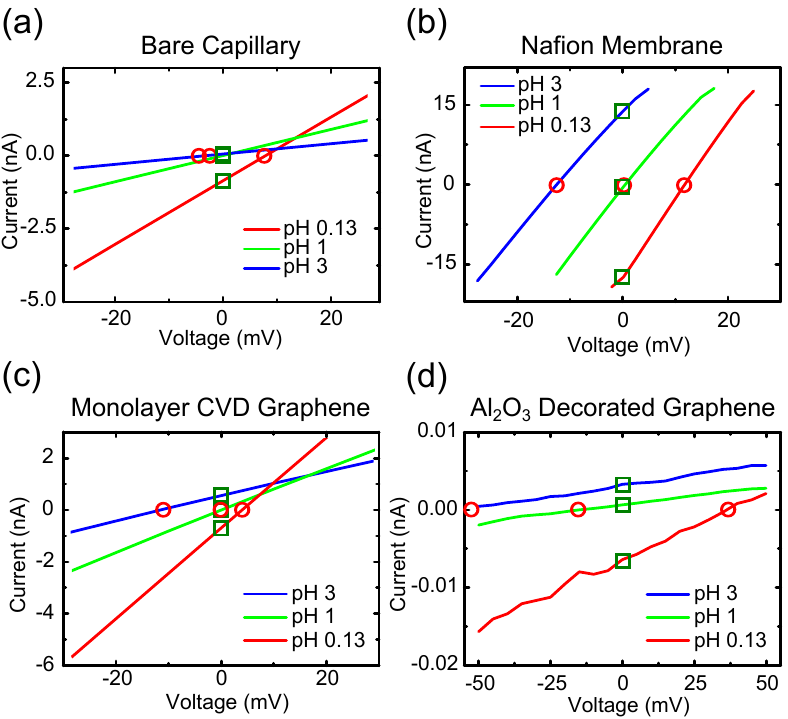}

\protect\caption{Typical I-V curves for different concentration differences. The concentration
in the capillary is 0.1~M HCl and the HCl solution in the reservoir
is indicated in the legends. The red circles indicate the reversal
potential (``A'') and the green squares indicate the diffusion driven
current (``B''). (a) For a bare capillary (180~nm diameter), the
lines pass close to 0,0 indicating minimal selectivity associated
with the charge on the glass. (b) The commercially available proton
selective membrane Nafion (2~\textgreek{m}m capillary). When there
is a concentration difference across the capillary a current flows
at 0V. This current can be stopped by applying the reversal potential.
(c) Typical I-V curves for an as grown monolayer graphene membrane,
small voltage offsets indicate that there is weak evidence for selectivity
(2~\textgreek{m}m capillary). (d) Typical I-V curves for a graphene
membrane decorated with Al\protect\textsubscript{2}O\protect\textsubscript{3}
(180~nm capillary). Here the voltage offsets are similar to those
for Nafion.\label{fig:Fig2}}
\end{figure}

Typical I-V curves for each material are shown in Fig~\ref{fig:Fig2}.
The I-V curves for a bare capillary show voltage offsets of less than
5~mV which correspond to a small component of the current being selective
due to the negative charge on the surface of the glass nanocapillaries.
A fully selective membrane would cause a reversal potential of 58~mV
per unit pH difference, so this indicates that less than 10\% of the
current is selective. Most of the current flows through the \foreignlanguage{british}{centre}
of the capillary and is carried equally by the positive and negative
ions. In contrast the Nafion membrane measurements show a clear shift
in the voltage and current as the concentration of HCl in the reservoir
changes (Fig~\ref{fig:Fig2}(b)). Offsets of 15 to 20~mV clearly
indicate that a significant proportion of the current is due to protons.
However, as the reversal potential does not shift by 58~mV it is
clear that there is also a significant leakage current. These results
confirm that our experiment can detect and quantify selectivity in
the presence of other ionic currents.

The I-V curves for an as grown graphene membrane show evidence for
limited selectivity, Fig~2(c). The voltage offsets are of the order
5~-~10~mV which indicates that a small proportion of the current
is proton selective. However, we find that the leakage currents dominate
over any selective current for our CVD graphene membranes.

Given that defects in CVD graphene are known to influence ionic flow
it is necessary to establish the extent to which defects affect the
current. We therefore used an atomic layer deposition process (ALD)
to block most of the defects. Al\textsubscript{2}O\textsubscript{3}
is deposited onto the NaCl supported graphene using a Cambridge Nanotech
Savannah ALD system with a 20~cycle process at 200\textdegree C,
consisting of alternating pulses of trimethylaluminium and water both
carried in a N\textsubscript{2}~(20~sccm) flow with 8~s purges
between them~\cite{Dlubak2012,Weatherup2013}. This method typically
yields a 2~nm thick film on Si with a native oxide. However, for
this relatively high-temperature, water-based process the poor wetting
of the Al\textsubscript{2}O\textsubscript{3} on graphene is well
documented, leading to preferential decoration at defects~\cite{Dlubak2012,Wang2008a,Robinson2010,Martin2014}.
It has been demonstrated that Al\textsubscript{2}O\textsubscript{3}
deposited by ALD reduces the ionic current~\cite{OHern2015} and
our samples indeed showed significantly increased resistance (shown
in supplementary information). Typical I-V curves for Al\textsubscript{2}O\textsubscript{3}
decorated graphene are shown in Fig~\ref{fig:Fig2}(d). The shape
of these I-V curves more closely resembles those observed for Nafion
than for either bare capillaries or as grown graphene. We measure
reversal potentials of 15~-~25~mV, indicating that a significant
proportion of the current is due to a proton flux, although the overall
currents are the lowest of the three.

\begin{figure*}
\includegraphics{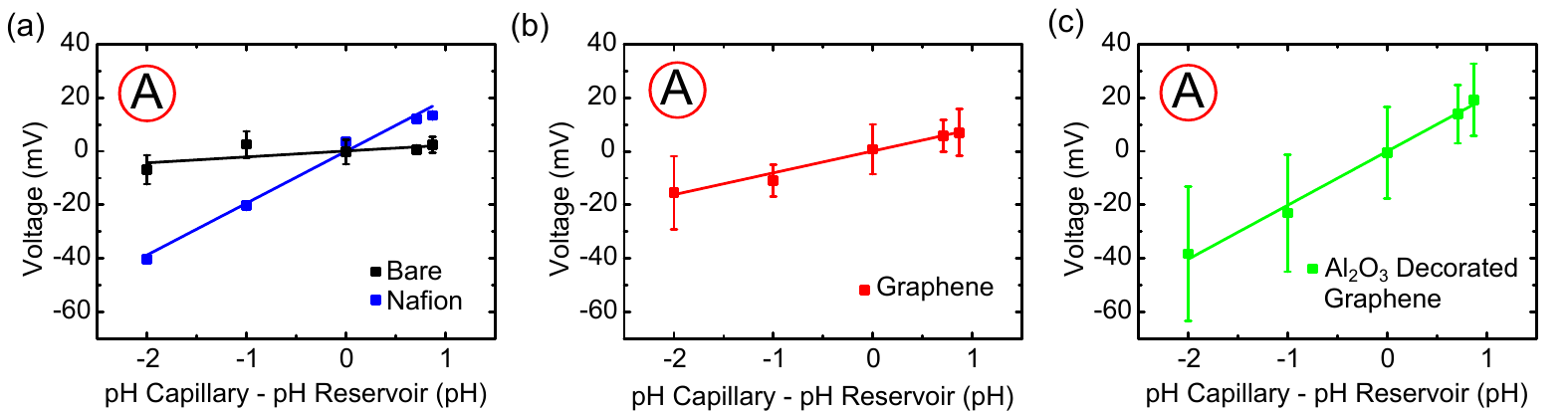}

\protect\caption{The voltage offsets plotted against the difference in capillary and
reservoir pH for each of the four conditions (this corresponds to
point ``A'' in Fig~\ref{fig:Fig1}(b)). (a) The bare capillaries
show a gradient of 2.2~mV/pH indicating the effect of negative charge
on the surfaces of the glass nanocapillaries. The Nafion shows a 19.4~mV/pH
offset indicating that the membrane is selective. (b) As grown graphene
has a voltage offset of 8.1~mV/pH. (c) Al\protect\textsubscript{2}O\protect\textsubscript{3}
decorated graphene shows an offset comparable with Nafion of 23~mV/pH
indicating significant selectivity. \label{fig:Fig3}}
\end{figure*}

We can further quantify the selectivity of the membranes by analysing
the reversal potentials (point A in Fig~\ref{fig:Fig1}(a)). These
are plotted in Fig~\ref{fig:Fig3}(a-c) against the difference in
pH of the solutions in the capillary and reservoir. From Fig~\ref{fig:Fig3}
we observe that none of the fitted lines have a gradient of 58~mV/pH
indicating that there is always a leakage current. The bare capillaries
show evidence of very low selectivity with a gradient of 2~mV/pH
whereas the Nafion is ten times more selective with a gradient of
20~mV/pH~(Fig~\ref{fig:Fig3}(a)). As grown graphene has a gradient
of 8~mV/pH, though the error bars are considerable~(Fig~\ref{fig:Fig3}(b)).
This means that whilst there is some selectivity the bulk of current
flow is due to non-selective leakage current. However, the graphene
decorated with Al\textsubscript{2}O\textsubscript{3} has a gradient
of 23~mV/pH (Fig~\ref{fig:Fig3}(c)) and exhibits proton selectivity
comparable to Nafion. 

The leakage conductivity for each membrane can be calculated from
the ratio of the observed voltage per pH to the expected voltage per
pH and the measured conductivity~\cite{Chimerel2013}. The leakage
and selective pathways form a voltage divider relating the measured
membrane voltage $V_{m}$, with the measured membrane conductance
$G_{m}$, and the leakage conductance $G_{0}$, to the Nernstian potential
$V$~\cite{Chimerel2013}.

\begin{equation}
V_{m}=V\frac{G_{m}-G_{0}}{G_{m}}\label{eq:MembraneVoltage}
\end{equation}

\begin{table*}
\begin{tabular}{|>{\centering}p{8cm}|>{\centering}p{2cm}|>{\centering}p{2cm}|>{\centering}p{2cm}|>{\centering}p{2cm}|}
\hline 
Membrane & Selectivity (mV/pH) & Conductivity nA/\textgreek{m}m\textsuperscript{2} & Leakage Conductivity nA/\textgreek{m}m\textsuperscript{2} & Selective H\textsuperscript{+} Conductivity nA/\textgreek{m}m\textsuperscript{2}\tabularnewline
\hline 
\hline 
Bare Glass Nanopore & $2.2\pm$0.56 & 800 & 740 & 60\tabularnewline
\hline 
Nafion &  $19.4\pm$0.78  & 18.7 & 12.5 & 6.2\tabularnewline
\hline 
\multicolumn{1}{|c|}{As grown CVD graphene} & $8.1\pm$0.70 & 16.4 & 15.1 & 1.3\tabularnewline
\hline 
Al\textsubscript{2}O\textsubscript{3} decorated graphene & $23\pm$1.6 & 1.2 & 0.8 & 0.4\tabularnewline
\hline 
\end{tabular}

\protect\caption{Table of average selectivity in mV/pH and the average conductivities
of different samples broken down in to leakage and selective current.\label{tab:Condutances}}
\end{table*}

Using eq~\ref{eq:MembraneVoltage} we can estimate the selective
proton conductivity for the as grown and Al\textsubscript{2}O\textsubscript{3}
decorated graphene, shown in Table~\ref{tab:Condutances}. 

\begin{figure}
\includegraphics{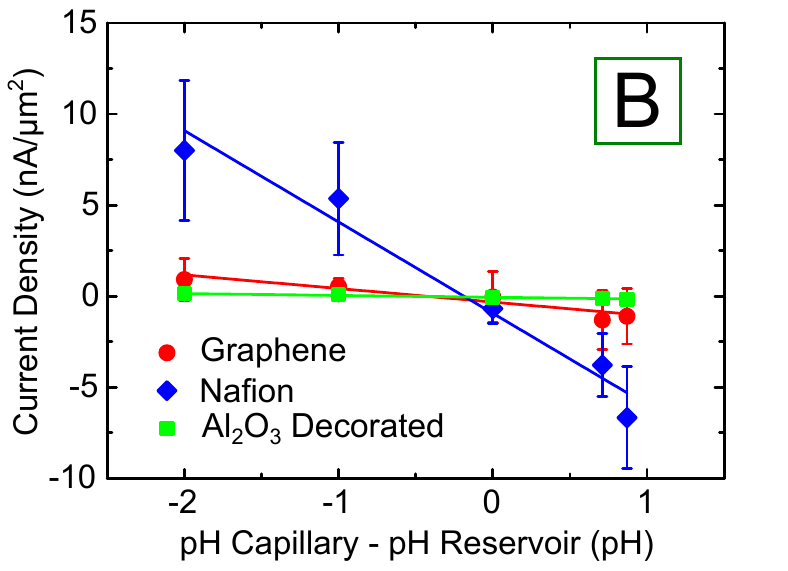}

\protect\caption{The current density at zero voltage plotted against the difference
in capillary and reservoir pH (this corresponds to point ``B'' in
Fig~\ref{fig:Fig1}(b)). This plot shows the concentration difference
driven current when the amplifier in voltage track mode holds the
voltage at 0V. All the samples show the current decreasing with increased
pH in the reservoir as expected for selectivity. However the Nafion
currents are much higher than those for graphene membranes. The lowest
currents are across the Al\protect\textsubscript{2}O\protect\textsubscript{3}
sample. These results are presented separately in the supplementary
information.\label{fig:CurrentOffsets}}

\end{figure}

To consider where this selectivity arises from it is instructive to
consider the current density at zero potential (point B in Fig~\ref{fig:Fig1}(b))
plotted in Fig~\ref{fig:CurrentOffsets}. For a selective membrane
the concentration gradient will drive a current when the potential
is zero. We see that both the as grown and decorated graphene show
a current that scales with pH. However for the Al\textsubscript{2}O\textsubscript{3}
decorated graphene it is much lower. We interpret this as current
flowing through defects in as grown graphene. Adding Al\textsubscript{2}O\textsubscript{3}
blocks the defects decreasing \emph{both} the selective and non selective
components of current. This can also be seen in the values of the
conductivities in Table~\ref{tab:Condutances}. The proton conductivity
for as grown graphene is 1.3~nA/\textgreek{m}m\textsuperscript{2}
compared to 0.4~nA/\textgreek{m}m\textsuperscript{2} for Al\textsubscript{2}O\textsubscript{3}
decorated graphene, despite the latter being three times more selective
to protons.

We propose that the increase in selectivity is due to the defects
decreasing in size and hence becoming more size selective for H\textsuperscript{+}
over Cl\textsuperscript{-}. Blocking the defects decreases both the
selective and non selective components of the conductivity through
the graphene. This shows that the small amount of selectivity observed
for the bare graphene membrane can also be attributed to selective
defects. If the protons passed directly through the membrane as opposed
to defects then we would not expect the proton current to decrease
as significantly when the defects are blocked. In contrast to a recently
published result~\cite{Hu2014}, this indicates that proton transport
across graphene membranes is via defects supporting the results in~\cite{Achtyl2015}
which comes to a similar conclusion for graphene supported on a substrate.

\begin{figure}
\includegraphics{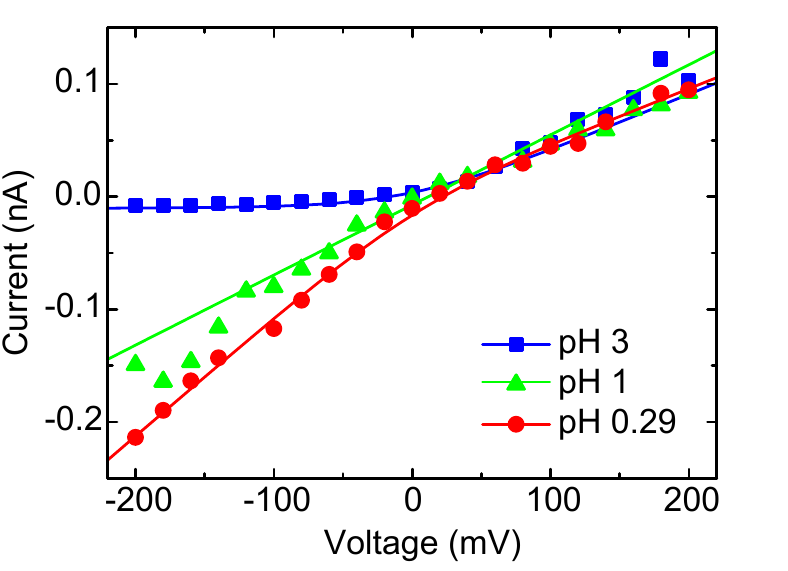}

\protect\caption{Plots of I-V curves for an Al\protect\textsubscript{2}O\protect\textsubscript{3}
decorated graphene membrane sealed across a 180~nm capillary. The
capillary contains 0.1~M HCl and the concentration in the reservoir
is exchanged. The lines show the fitted GHK equation, from which the
permeability coefficients can be be determined.\label{fig:(4)}}
\end{figure}

An alternative analysis of our I-V characteristics can also be made
by considering the Goldman Hodgkin Katz (GHK) equations~\cite{Goldman1943}.
These predict the shape of the I-V curve expected for a selective
membrane~\cite{Hille2001}. See supplementary information for further
details. We only observed results which correspond to the GHK equations
for the Al\textsubscript{2}O\textsubscript{3} decorated graphene
membrane (Fig~\ref{fig:(4)}). When the capillary is at a higher
concentration than the reservoir, positive currents are larger indicating
that protons can cross from the capillary into the reservoir. However
when a negative voltage is applied the current is decreased since
there are fewer protons in the reservoir to cross into the capillary,
demonstrating that the Cl\textsuperscript{-} ions in excess in the
capillary are unable to cross the membrane and contribute to the current
(blue line Fig~\ref{fig:(4)}). When the concentration gradient is
reversed the relative currents switch around so that the current due
to a negative applied voltage is higher than the current for a positive
applied voltage (red line Fig~\ref{fig:(4)}). The ratio between
the permeability of the positive and negative ion indicates proton
selectivity. When the reservoir is at pH~0.29 we measure a ratio
of 3.5 which would correspond to a reversal potential of 20~mV, consistent
with the reversal potentials we have observed.

The high relative selectivity of the Al\textsubscript{2}O\textsubscript{3}
decorated graphene suggests it could be \foreignlanguage{british}{utilised}
as a selective membrane, for example in fuel cells. However, on the
basis of our results we think there are significant challenges as
the proton fluxes observed are much lower than for Nafion. The graphene
membranes are either less selective or less permeable to protons than
their commercial competitors.

Our results show that defects are critical for ionic transport and
selectivity in our graphene membranes. Understanding the nature of
these defects and how their size and chemistry influences their selectivity
is a key challenge. We have shown that even a very small number of
defects, or partially blocked defects can dominate ionic transport
properties. Verifying that no defects exist in a graphene membrane
is extremely challenging. The minimum defect density detectable in
Raman spectroscopy is about $2\times10^{9}cm^{-2}$ hence up to 20
defects may exist per 1~\textgreek{m}m\textsuperscript{2} of graphene
where Raman spectroscopy indicates defect free graphene~\cite{Cancado2011,Ferrari2013,Lucchese2010}.
Conversely TEM or STM have sufficient resolution to image individual
defects but it is not practical to image the \foreignlanguage{british}{entirety}
of the graphene covering a 2~\textgreek{m}m pore.

We have measured the selectivity of Nafion and graphene membranes
to H\textsuperscript{+} by analysing the I-V characteristics when
there is a concentration difference across the membrane. Our results
show that a leakage current does not prevent the use of established
techniques to probe selectivity in aqueous solutions. By studying
reversal potentials we have identified a small proton selective current
through graphene membranes which can be reduced by blocking defects.
We conclude that the proton selective current is through defects as
opposed to across the graphene membrane. 

\begin{acknowledgments}
The authors would like to thank C. Chimeral for useful discussions. This work was supported by the EPSRC Cambridge NanoDTC, EP/G037221/1 and EPSRC grant GRAPHTED, EP/K016636/1. RSW acknowledges a Research Fellowship from St. John's College, Cambridge. All data accompanying this publication are directly available within the publication.
\end{acknowledgments}

\bibliographystyle{apsrev4-1}
\bibliography{ProtonPaperDraft1A}

\begin{thebibliography}{26}%
\makeatletter
\providecommand \@ifxundefined [1]{%
 \@ifx{#1\undefined}
}%
\providecommand \@ifnum [1]{%
 \ifnum #1\expandafter \@firstoftwo
 \else \expandafter \@secondoftwo
 \fi
}%
\providecommand \@ifx [1]{%
 \ifx #1\expandafter \@firstoftwo
 \else \expandafter \@secondoftwo
 \fi
}%
\providecommand \natexlab [1]{#1}%
\providecommand \enquote  [1]{``#1''}%
\providecommand \bibnamefont  [1]{#1}%
\providecommand \bibfnamefont [1]{#1}%
\providecommand \citenamefont [1]{#1}%
\providecommand \href@noop [0]{\@secondoftwo}%
\providecommand \href [0]{\begingroup \@sanitize@url \@href}%
\providecommand \@href[1]{\@@startlink{#1}\@@href}%
\providecommand \@@href[1]{\endgroup#1\@@endlink}%
\providecommand \@sanitize@url [0]{\catcode `\\12\catcode `\$12\catcode
  `\&12\catcode `\#12\catcode `\^12\catcode `\_12\catcode `\%12\relax}%
\providecommand \@@startlink[1]{}%
\providecommand \@@endlink[0]{}%
\providecommand \url  [0]{\begingroup\@sanitize@url \@url }%
\providecommand \@url [1]{\endgroup\@href {#1}{\urlprefix }}%
\providecommand \urlprefix  [0]{URL }%
\providecommand \Eprint [0]{\href }%
\providecommand \doibase [0]{http://dx.doi.org/}%
\providecommand \selectlanguage [0]{\@gobble}%
\providecommand \bibinfo  [0]{\@secondoftwo}%
\providecommand \bibfield  [0]{\@secondoftwo}%
\providecommand \translation [1]{[#1]}%
\providecommand \BibitemOpen [0]{}%
\providecommand \bibitemStop [0]{}%
\providecommand \bibitemNoStop [0]{.\EOS\space}%
\providecommand \EOS [0]{\spacefactor3000\relax}%
\providecommand \BibitemShut  [1]{\csname bibitem#1\endcsname}%
\let\auto@bib@innerbib\@empty
%</preamble>
\bibitem [{\citenamefont {Sint}\ \emph {et~al.}(2008)\citenamefont {Sint},
  \citenamefont {Wang}, \citenamefont {Kr\'{a}l},\ and\ \citenamefont
  {Kra}}]{Sint2008}%
  \BibitemOpen
  \bibfield  {author} {\bibinfo {author} {\bibfnamefont {K.}~\bibnamefont
  {Sint}}, \bibinfo {author} {\bibfnamefont {B.}~\bibnamefont {Wang}}, \bibinfo
  {author} {\bibfnamefont {P.}~\bibnamefont {Kr\'{a}l}}, \ and\ \bibinfo
  {author} {\bibfnamefont {P.}~\bibnamefont {Kra}},\ }\href {\doibase
  10.1021/ja804409f} {\bibfield  {journal} {\bibinfo  {journal} {J. Am. Chem.
  Soc.}\ }\textbf {\bibinfo {volume} {130}},\ \bibinfo {pages} {16448}
  (\bibinfo {year} {2008})}\BibitemShut {NoStop}%
\bibitem [{\citenamefont {Kang}\ \emph {et~al.}(2014)\citenamefont {Kang},
  \citenamefont {Zhang}, \citenamefont {Shi}, \citenamefont {Zhang},
  \citenamefont {Liang}, \citenamefont {Wang}, \citenamefont {Agren},\ and\
  \citenamefont {Tu}}]{Kang2014}%
  \BibitemOpen
  \bibfield  {author} {\bibinfo {author} {\bibfnamefont {Y.}~\bibnamefont
  {Kang}}, \bibinfo {author} {\bibfnamefont {Z.}~\bibnamefont {Zhang}},
  \bibinfo {author} {\bibfnamefont {H.}~\bibnamefont {Shi}}, \bibinfo {author}
  {\bibfnamefont {J.}~\bibnamefont {Zhang}}, \bibinfo {author} {\bibfnamefont
  {L.}~\bibnamefont {Liang}}, \bibinfo {author} {\bibfnamefont
  {Q.}~\bibnamefont {Wang}}, \bibinfo {author} {\bibfnamefont {H.}~\bibnamefont
  {Agren}}, \ and\ \bibinfo {author} {\bibfnamefont {Y.}~\bibnamefont {Tu}},\
  }\href {\doibase 10.1039/c4nr01383b} {\bibfield  {journal} {\bibinfo
  {journal} {Nanoscale}\ }\textbf {\bibinfo {volume} {6}},\ \bibinfo {pages}
  {10666} (\bibinfo {year} {2014})}\BibitemShut {NoStop}%
\bibitem [{\citenamefont {Surwade}\ \emph {et~al.}(2015)\citenamefont
  {Surwade}, \citenamefont {Smirnov}, \citenamefont {Vlassiouk}, \citenamefont
  {Unocic}, \citenamefont {Veith}, \citenamefont {Dai},\ and\ \citenamefont
  {Mahurin}}]{Surwade2015}%
  \BibitemOpen
  \bibfield  {author} {\bibinfo {author} {\bibfnamefont {S.~P.}\ \bibnamefont
  {Surwade}}, \bibinfo {author} {\bibfnamefont {S.~N.}\ \bibnamefont
  {Smirnov}}, \bibinfo {author} {\bibfnamefont {I.~V.}\ \bibnamefont
  {Vlassiouk}}, \bibinfo {author} {\bibfnamefont {R.~R.}\ \bibnamefont
  {Unocic}}, \bibinfo {author} {\bibfnamefont {G.~M.}\ \bibnamefont {Veith}},
  \bibinfo {author} {\bibfnamefont {S.}~\bibnamefont {Dai}}, \ and\ \bibinfo
  {author} {\bibfnamefont {S.~M.}\ \bibnamefont {Mahurin}},\ }\href {\doibase
  10.1038/nnano.2015.37} {\bibfield  {journal} {\bibinfo  {journal} {Nature
  Nanotechnology}\ ,\ \bibinfo {pages} {1}} (\bibinfo {year}
  {2015})}\BibitemShut {NoStop}%
\bibitem [{\citenamefont {O'Hern}\ \emph {et~al.}(2012)\citenamefont {O'Hern},
  \citenamefont {Stewart}, \citenamefont {Boutilier}, \citenamefont {Idrobo},
  \citenamefont {Bhaviripudi}, \citenamefont {Das}, \citenamefont {Kong},
  \citenamefont {Laoui}, \citenamefont {Atieh},\ and\ \citenamefont
  {Karnik}}]{O'Hern2012}%
  \BibitemOpen
  \bibfield  {author} {\bibinfo {author} {\bibfnamefont {S.~C.}\ \bibnamefont
  {O'Hern}}, \bibinfo {author} {\bibfnamefont {C.~A.}\ \bibnamefont {Stewart}},
  \bibinfo {author} {\bibfnamefont {M.~S.~H.}\ \bibnamefont {Boutilier}},
  \bibinfo {author} {\bibfnamefont {J.-C.}\ \bibnamefont {Idrobo}}, \bibinfo
  {author} {\bibfnamefont {S.}~\bibnamefont {Bhaviripudi}}, \bibinfo {author}
  {\bibfnamefont {S.~K.}\ \bibnamefont {Das}}, \bibinfo {author} {\bibfnamefont
  {J.}~\bibnamefont {Kong}}, \bibinfo {author} {\bibfnamefont {T.}~\bibnamefont
  {Laoui}}, \bibinfo {author} {\bibfnamefont {M.}~\bibnamefont {Atieh}}, \ and\
  \bibinfo {author} {\bibfnamefont {R.}~\bibnamefont {Karnik}},\ }\href
  {\doibase 10.1021/nn303869m} {\bibfield  {journal} {\bibinfo  {journal} {ACS
  Nano}\ }\textbf {\bibinfo {volume} {6}},\ \bibinfo {pages} {10130} (\bibinfo
  {year} {2012})}\BibitemShut {NoStop}%
\bibitem [{\citenamefont {O'Hern}\ \emph {et~al.}(2014)\citenamefont {O'Hern},
  \citenamefont {Boutilier}, \citenamefont {Idrobo}, \citenamefont {Song},
  \citenamefont {Kong}, \citenamefont {Laoui}, \citenamefont {Atieh},\ and\
  \citenamefont {Karnik}}]{O'Hern2014}%
  \BibitemOpen
  \bibfield  {author} {\bibinfo {author} {\bibfnamefont {S.~C.}\ \bibnamefont
  {O'Hern}}, \bibinfo {author} {\bibfnamefont {M.~S.~H.}\ \bibnamefont
  {Boutilier}}, \bibinfo {author} {\bibfnamefont {J.-C.~C.}\ \bibnamefont
  {Idrobo}}, \bibinfo {author} {\bibfnamefont {Y.}~\bibnamefont {Song}},
  \bibinfo {author} {\bibfnamefont {J.}~\bibnamefont {Kong}}, \bibinfo {author}
  {\bibfnamefont {T.}~\bibnamefont {Laoui}}, \bibinfo {author} {\bibfnamefont
  {M.}~\bibnamefont {Atieh}}, \ and\ \bibinfo {author} {\bibfnamefont
  {R.}~\bibnamefont {Karnik}},\ }\href {\doibase 10.1021/nl404118f} {\bibfield
  {journal} {\bibinfo  {journal} {Nano Letters}\ }\textbf {\bibinfo {volume}
  {14}},\ \bibinfo {pages} {1234 } (\bibinfo {year} {2014})}\BibitemShut
  {NoStop}%
\bibitem [{\citenamefont {Koenig}\ \emph {et~al.}(2012)\citenamefont {Koenig},
  \citenamefont {Wang}, \citenamefont {Pellegrino},\ and\ \citenamefont
  {Bunch}}]{Koenig2012}%
  \BibitemOpen
  \bibfield  {author} {\bibinfo {author} {\bibfnamefont {S.~P.}\ \bibnamefont
  {Koenig}}, \bibinfo {author} {\bibfnamefont {L.}~\bibnamefont {Wang}},
  \bibinfo {author} {\bibfnamefont {J.}~\bibnamefont {Pellegrino}}, \ and\
  \bibinfo {author} {\bibfnamefont {J.~S.}\ \bibnamefont {Bunch}},\ }\href
  {\doibase 10.1038/nnano.2012.162} {\bibfield  {journal} {\bibinfo  {journal}
  {Nature Nanotechnology}\ }\textbf {\bibinfo {volume} {7}},\ \bibinfo {pages}
  {728} (\bibinfo {year} {2012})}\BibitemShut {NoStop}%
\bibitem [{\citenamefont {Garaj}\ \emph {et~al.}(2010)\citenamefont {Garaj},
  \citenamefont {Hubbard}, \citenamefont {Reina}, \citenamefont {Kong},
  \citenamefont {Branton},\ and\ \citenamefont {Golovchenko}}]{Garaj2010a}%
  \BibitemOpen
  \bibfield  {author} {\bibinfo {author} {\bibfnamefont {S.}~\bibnamefont
  {Garaj}}, \bibinfo {author} {\bibfnamefont {W.}~\bibnamefont {Hubbard}},
  \bibinfo {author} {\bibfnamefont {A.}~\bibnamefont {Reina}}, \bibinfo
  {author} {\bibfnamefont {J.}~\bibnamefont {Kong}}, \bibinfo {author}
  {\bibfnamefont {D.}~\bibnamefont {Branton}}, \ and\ \bibinfo {author}
  {\bibfnamefont {J.~A.}\ \bibnamefont {Golovchenko}},\ }\href {\doibase
  10.1038/nature09379} {\bibfield  {journal} {\bibinfo  {journal} {Nature}\
  }\textbf {\bibinfo {volume} {467}},\ \bibinfo {pages} {190} (\bibinfo {year}
  {2010})}\BibitemShut {NoStop}%
\bibitem [{\citenamefont {Schneider}\ \emph {et~al.}(2010)\citenamefont
  {Schneider}, \citenamefont {Kowalczyk}, \citenamefont {Calado}, \citenamefont
  {Pandraud}, \citenamefont {Zandbergen}, \citenamefont {Vandersypen},\ and\
  \citenamefont {Dekker}}]{Schneider2010}%
  \BibitemOpen
  \bibfield  {author} {\bibinfo {author} {\bibfnamefont {G.~F.}\ \bibnamefont
  {Schneider}}, \bibinfo {author} {\bibfnamefont {S.~W.}\ \bibnamefont
  {Kowalczyk}}, \bibinfo {author} {\bibfnamefont {V.~E.}\ \bibnamefont
  {Calado}}, \bibinfo {author} {\bibfnamefont {G.}~\bibnamefont {Pandraud}},
  \bibinfo {author} {\bibfnamefont {H.~W.}\ \bibnamefont {Zandbergen}},
  \bibinfo {author} {\bibfnamefont {L.~M.~K.}\ \bibnamefont {Vandersypen}}, \
  and\ \bibinfo {author} {\bibfnamefont {C.}~\bibnamefont {Dekker}},\ }\href
  {\doibase 10.1021/nl102069z} {\bibfield  {journal} {\bibinfo  {journal} {Nano
  Letters}\ }\textbf {\bibinfo {volume} {10}},\ \bibinfo {pages} {3163}
  (\bibinfo {year} {2010})}\BibitemShut {NoStop}%
\bibitem [{\citenamefont {Merchant}\ \emph {et~al.}(2010)\citenamefont
  {Merchant}, \citenamefont {Healy}, \citenamefont {Wanunu}, \citenamefont
  {Ray}, \citenamefont {Peterman}, \citenamefont {Bartel}, \citenamefont
  {Fischbein}, \citenamefont {Venta}, \citenamefont {Luo}, \citenamefont
  {Johnson},\ and\ \citenamefont {Drndi\'{c}}}]{Merchant2010}%
  \BibitemOpen
  \bibfield  {author} {\bibinfo {author} {\bibfnamefont {C.~A.}\ \bibnamefont
  {Merchant}}, \bibinfo {author} {\bibfnamefont {K.}~\bibnamefont {Healy}},
  \bibinfo {author} {\bibfnamefont {M.}~\bibnamefont {Wanunu}}, \bibinfo
  {author} {\bibfnamefont {V.}~\bibnamefont {Ray}}, \bibinfo {author}
  {\bibfnamefont {N.}~\bibnamefont {Peterman}}, \bibinfo {author}
  {\bibfnamefont {J.}~\bibnamefont {Bartel}}, \bibinfo {author} {\bibfnamefont
  {M.~D.}\ \bibnamefont {Fischbein}}, \bibinfo {author} {\bibfnamefont
  {K.}~\bibnamefont {Venta}}, \bibinfo {author} {\bibfnamefont
  {Z.}~\bibnamefont {Luo}}, \bibinfo {author} {\bibfnamefont {A.~T.~C.}\
  \bibnamefont {Johnson}}, \ and\ \bibinfo {author} {\bibfnamefont
  {M.}~\bibnamefont {Drndi\'{c}}},\ }\href {\doibase 10.1021/nl101046t}
  {\bibfield  {journal} {\bibinfo  {journal} {Nano Letters}\ }\textbf {\bibinfo
  {volume} {10}},\ \bibinfo {pages} {2915} (\bibinfo {year}
  {2010})}\BibitemShut {NoStop}%
\bibitem [{\citenamefont {Hu}\ \emph {et~al.}(2014)\citenamefont {Hu},
  \citenamefont {Lozada-Hidalgo}, \citenamefont {Wang}, \citenamefont
  {Mishchenko}, \citenamefont {Schedin}, \citenamefont {Nair}, \citenamefont
  {Hill}, \citenamefont {Boukhvalov}, \citenamefont {Katsnelson}, \citenamefont
  {Dryfe}, \citenamefont {Grigorieva}, \citenamefont {Wu},\ and\ \citenamefont
  {Geim}}]{Hu2014}%
  \BibitemOpen
  \bibfield  {author} {\bibinfo {author} {\bibfnamefont {S.}~\bibnamefont
  {Hu}}, \bibinfo {author} {\bibfnamefont {M.}~\bibnamefont {Lozada-Hidalgo}},
  \bibinfo {author} {\bibfnamefont {F.~C.}\ \bibnamefont {Wang}}, \bibinfo
  {author} {\bibfnamefont {A.}~\bibnamefont {Mishchenko}}, \bibinfo {author}
  {\bibfnamefont {F.}~\bibnamefont {Schedin}}, \bibinfo {author} {\bibfnamefont
  {R.~R.}\ \bibnamefont {Nair}}, \bibinfo {author} {\bibfnamefont {E.~W.}\
  \bibnamefont {Hill}}, \bibinfo {author} {\bibfnamefont {D.~W.}\ \bibnamefont
  {Boukhvalov}}, \bibinfo {author} {\bibfnamefont {M.~I.}\ \bibnamefont
  {Katsnelson}}, \bibinfo {author} {\bibfnamefont {R.~A.~W.}\ \bibnamefont
  {Dryfe}}, \bibinfo {author} {\bibfnamefont {I.~V.}\ \bibnamefont
  {Grigorieva}}, \bibinfo {author} {\bibfnamefont {H.~A.}\ \bibnamefont {Wu}},
  \ and\ \bibinfo {author} {\bibfnamefont {A.~K.}\ \bibnamefont {Geim}},\
  }\href {\doibase 10.1038/nature14015} {\bibfield  {journal} {\bibinfo
  {journal} {Nature}\ }\textbf {\bibinfo {volume} {516}},\ \bibinfo {pages}
  {227} (\bibinfo {year} {2014})}\BibitemShut {NoStop}%
\bibitem [{\citenamefont {Hille}(2001)}]{Hille2001}%
  \BibitemOpen
  \bibfield  {author} {\bibinfo {author} {\bibfnamefont {B.}~\bibnamefont
  {Hille}},\ }\href@noop {} {\emph {\bibinfo {title} {{Ion Channels of
  Excitable Membranes}}}},\ \bibinfo {edition} {3rd}\ ed.\ (\bibinfo
  {publisher} {Sinauer Associates},\ \bibinfo {year} {2001})\ pp.\ \bibinfo
  {pages} {441 -- 470}\BibitemShut {NoStop}%
\bibitem [{\citenamefont {Walker}\ \emph {et~al.}(2015)\citenamefont {Walker},
  \citenamefont {Weatherup}, \citenamefont {Bell}, \citenamefont {Hofmann},\
  and\ \citenamefont {Keyser}}]{Walker2015}%
  \BibitemOpen
  \bibfield  {author} {\bibinfo {author} {\bibfnamefont {M.~I.}\ \bibnamefont
  {Walker}}, \bibinfo {author} {\bibfnamefont {R.~S.}\ \bibnamefont
  {Weatherup}}, \bibinfo {author} {\bibfnamefont {N.~A.~W.}\ \bibnamefont
  {Bell}}, \bibinfo {author} {\bibfnamefont {S.}~\bibnamefont {Hofmann}}, \
  and\ \bibinfo {author} {\bibfnamefont {U.~F.}\ \bibnamefont {Keyser}},\
  }\href {\doibase 10.1063/1.4906236} {\bibfield  {journal} {\bibinfo
  {journal} {Applied Physics Letters}\ }\textbf {\bibinfo {volume} {106}},\
  \bibinfo {pages} {023119} (\bibinfo {year} {2015})}\BibitemShut {NoStop}%
\bibitem [{\citenamefont {Hofmann}\ \emph {et~al.}(2015)\citenamefont
  {Hofmann}, \citenamefont {Braeuninger-Weimer},\ and\ \citenamefont
  {Weatherup}}]{Hofmann2015}%
  \BibitemOpen
  \bibfield  {author} {\bibinfo {author} {\bibfnamefont {S.}~\bibnamefont
  {Hofmann}}, \bibinfo {author} {\bibfnamefont {P.}~\bibnamefont
  {Braeuninger-Weimer}}, \ and\ \bibinfo {author} {\bibfnamefont {R.~S.}\
  \bibnamefont {Weatherup}},\ }\href {\doibase 10.1021/acs.jpclett.5b01052}
  {\bibfield  {journal} {\bibinfo  {journal} {The Journal of Physical Chemistry
  Letters}\ }\textbf {\bibinfo {volume} {6}},\ \bibinfo {pages} {2714}
  (\bibinfo {year} {2015})}\BibitemShut {NoStop}%
\bibitem [{\citenamefont {Mauritz}\ and\ \citenamefont
  {Moore}(2004)}]{Mauritz2004}%
  \BibitemOpen
  \bibfield  {author} {\bibinfo {author} {\bibfnamefont {K.~A.}\ \bibnamefont
  {Mauritz}}\ and\ \bibinfo {author} {\bibfnamefont {R.~B.}\ \bibnamefont
  {Moore}},\ }\href {\doibase 10.1021/cr0207123} {\bibfield  {journal}
  {\bibinfo  {journal} {Chemical Reviews}\ }\textbf {\bibinfo {volume} {104}},\
  \bibinfo {pages} {4535} (\bibinfo {year} {2004})}\BibitemShut {NoStop}%
\bibitem [{\citenamefont {Dlubak}\ \emph {et~al.}(2012)\citenamefont {Dlubak},
  \citenamefont {Kidambi}, \citenamefont {Weatherup}, \citenamefont {Hofmann},\
  and\ \citenamefont {Robertson}}]{Dlubak2012}%
  \BibitemOpen
  \bibfield  {author} {\bibinfo {author} {\bibfnamefont {B.}~\bibnamefont
  {Dlubak}}, \bibinfo {author} {\bibfnamefont {P.~R.}\ \bibnamefont {Kidambi}},
  \bibinfo {author} {\bibfnamefont {R.~S.}\ \bibnamefont {Weatherup}}, \bibinfo
  {author} {\bibfnamefont {S.}~\bibnamefont {Hofmann}}, \ and\ \bibinfo
  {author} {\bibfnamefont {J.}~\bibnamefont {Robertson}},\ }\href {\doibase
  10.1063/1.4707376} {\bibfield  {journal} {\bibinfo  {journal} {Applied
  Physics Letters}\ }\textbf {\bibinfo {volume} {100}},\ \bibinfo {pages}
  {173113} (\bibinfo {year} {2012})}\BibitemShut {NoStop}%
\bibitem [{\citenamefont {Weatherup}\ \emph {et~al.}(2013)\citenamefont
  {Weatherup}, \citenamefont {Baehtz}, \citenamefont {Dlubak}, \citenamefont
  {Bayer}, \citenamefont {Kidambi}, \citenamefont {Blume}, \citenamefont
  {Schloegl},\ and\ \citenamefont {Hofmann}}]{Weatherup2013}%
  \BibitemOpen
  \bibfield  {author} {\bibinfo {author} {\bibfnamefont {R.~S.}\ \bibnamefont
  {Weatherup}}, \bibinfo {author} {\bibfnamefont {C.}~\bibnamefont {Baehtz}},
  \bibinfo {author} {\bibfnamefont {B.}~\bibnamefont {Dlubak}}, \bibinfo
  {author} {\bibfnamefont {B.~C.}\ \bibnamefont {Bayer}}, \bibinfo {author}
  {\bibfnamefont {P.~R.}\ \bibnamefont {Kidambi}}, \bibinfo {author}
  {\bibfnamefont {R.}~\bibnamefont {Blume}}, \bibinfo {author} {\bibfnamefont
  {R.}~\bibnamefont {Schloegl}}, \ and\ \bibinfo {author} {\bibfnamefont
  {S.}~\bibnamefont {Hofmann}},\ }\href@noop {} {\bibfield  {journal} {\bibinfo
   {journal} {Nano Letters}\ }\textbf {\bibinfo {volume} {13}},\ \bibinfo
  {pages} {4624} (\bibinfo {year} {2013})}\BibitemShut {NoStop}%
\bibitem [{\citenamefont {Wang}\ \emph {et~al.}(2008)\citenamefont {Wang},
  \citenamefont {Tabakman},\ and\ \citenamefont {Dai}}]{Wang2008a}%
  \BibitemOpen
  \bibfield  {author} {\bibinfo {author} {\bibfnamefont {X.}~\bibnamefont
  {Wang}}, \bibinfo {author} {\bibfnamefont {S.}~\bibnamefont {Tabakman}}, \
  and\ \bibinfo {author} {\bibfnamefont {H.}~\bibnamefont {Dai}},\ }\href
  {\doibase 10.1021/ja8023059} {\bibfield  {journal} {\bibinfo  {journal} {J.
  Am. Chem. Soc.}\ }\textbf {\bibinfo {volume} {130}},\ \bibinfo {pages} {8152}
  (\bibinfo {year} {2008})}\BibitemShut {NoStop}%
\bibitem [{\citenamefont {Robinson}\ \emph {et~al.}(2010)\citenamefont
  {Robinson}, \citenamefont {Labella}, \citenamefont {Trumbull}, \citenamefont
  {Weng}, \citenamefont {Cavelero}, \citenamefont {Daniels}, \citenamefont
  {Hughes}, \citenamefont {Hollander}, \citenamefont {Fanton},\ and\
  \citenamefont {Snyder}}]{Robinson2010}%
  \BibitemOpen
  \bibfield  {author} {\bibinfo {author} {\bibfnamefont {J.~A.}\ \bibnamefont
  {Robinson}}, \bibinfo {author} {\bibfnamefont {M.}~\bibnamefont {Labella}},
  \bibinfo {author} {\bibfnamefont {K.~A.}\ \bibnamefont {Trumbull}}, \bibinfo
  {author} {\bibfnamefont {X.}~\bibnamefont {Weng}}, \bibinfo {author}
  {\bibfnamefont {R.}~\bibnamefont {Cavelero}}, \bibinfo {author}
  {\bibfnamefont {T.}~\bibnamefont {Daniels}}, \bibinfo {author} {\bibfnamefont
  {Z.}~\bibnamefont {Hughes}}, \bibinfo {author} {\bibfnamefont
  {M.}~\bibnamefont {Hollander}}, \bibinfo {author} {\bibfnamefont
  {M.}~\bibnamefont {Fanton}}, \ and\ \bibinfo {author} {\bibfnamefont
  {D.}~\bibnamefont {Snyder}},\ }\href {\doibase 10.1021/nn1003138} {\bibfield
  {journal} {\bibinfo  {journal} {ACS Nano}\ }\textbf {\bibinfo {volume} {4}},\
  \bibinfo {pages} {2667} (\bibinfo {year} {2010})}\BibitemShut {NoStop}%
\bibitem [{\citenamefont {Martin}\ \emph {et~al.}(2014)\citenamefont {Martin},
  \citenamefont {Dlubak}, \citenamefont {Weatherup}, \citenamefont {Yang},
  \citenamefont {Deranlot}, \citenamefont {Bouzehouane}, \citenamefont
  {Petroff}, \citenamefont {Anane}, \citenamefont {Hofmann}, \citenamefont
  {Robertson}, \citenamefont {Fert},\ and\ \citenamefont
  {Seneor}}]{Martin2014}%
  \BibitemOpen
  \bibfield  {author} {\bibinfo {author} {\bibfnamefont {M.~B.}\ \bibnamefont
  {Martin}}, \bibinfo {author} {\bibfnamefont {B.}~\bibnamefont {Dlubak}},
  \bibinfo {author} {\bibfnamefont {R.~S.}\ \bibnamefont {Weatherup}}, \bibinfo
  {author} {\bibfnamefont {H.}~\bibnamefont {Yang}}, \bibinfo {author}
  {\bibfnamefont {C.}~\bibnamefont {Deranlot}}, \bibinfo {author}
  {\bibfnamefont {K.}~\bibnamefont {Bouzehouane}}, \bibinfo {author}
  {\bibfnamefont {F.}~\bibnamefont {Petroff}}, \bibinfo {author} {\bibfnamefont
  {A.}~\bibnamefont {Anane}}, \bibinfo {author} {\bibfnamefont
  {S.}~\bibnamefont {Hofmann}}, \bibinfo {author} {\bibfnamefont
  {J.}~\bibnamefont {Robertson}}, \bibinfo {author} {\bibfnamefont
  {A.}~\bibnamefont {Fert}}, \ and\ \bibinfo {author} {\bibfnamefont
  {P.}~\bibnamefont {Seneor}},\ }\href@noop {} {\bibfield  {journal} {\bibinfo
  {journal} {ACS Nano}\ }\textbf {\bibinfo {volume} {8}},\ \bibinfo {pages}
  {7890} (\bibinfo {year} {2014})}\BibitemShut {NoStop}%
\bibitem [{\citenamefont {O'Hern}\ \emph {et~al.}(2015)\citenamefont {O'Hern},
  \citenamefont {Jang}, \citenamefont {Bose}, \citenamefont {Idrobo},
  \citenamefont {Song}, \citenamefont {Laoui}, \citenamefont {Kong},\ and\
  \citenamefont {Karnik}}]{OHern2015}%
  \BibitemOpen
  \bibfield  {author} {\bibinfo {author} {\bibfnamefont {S.~C.}\ \bibnamefont
  {O'Hern}}, \bibinfo {author} {\bibfnamefont {D.}~\bibnamefont {Jang}},
  \bibinfo {author} {\bibfnamefont {S.}~\bibnamefont {Bose}}, \bibinfo {author}
  {\bibfnamefont {J.-C.}\ \bibnamefont {Idrobo}}, \bibinfo {author}
  {\bibfnamefont {Y.}~\bibnamefont {Song}}, \bibinfo {author} {\bibfnamefont
  {T.}~\bibnamefont {Laoui}}, \bibinfo {author} {\bibfnamefont
  {J.}~\bibnamefont {Kong}}, \ and\ \bibinfo {author} {\bibfnamefont
  {R.}~\bibnamefont {Karnik}},\ }\href {\doibase 10.1021/acs.nanolett.5b00456}
  {\bibfield  {journal} {\bibinfo  {journal} {Nano Letters}\ }\textbf {\bibinfo
  {volume} {15}},\ \bibinfo {pages} {3254} (\bibinfo {year}
  {2015})}\BibitemShut {NoStop}%
\bibitem [{\citenamefont {Chimerel}\ \emph {et~al.}(2013)\citenamefont
  {Chimerel}, \citenamefont {Murray}, \citenamefont {Oldewurtel}, \citenamefont
  {Summers},\ and\ \citenamefont {Keyser}}]{Chimerel2013}%
  \BibitemOpen
  \bibfield  {author} {\bibinfo {author} {\bibfnamefont {C.}~\bibnamefont
  {Chimerel}}, \bibinfo {author} {\bibfnamefont {A.~J.}\ \bibnamefont
  {Murray}}, \bibinfo {author} {\bibfnamefont {E.~R.}\ \bibnamefont
  {Oldewurtel}}, \bibinfo {author} {\bibfnamefont {D.~K.}\ \bibnamefont
  {Summers}}, \ and\ \bibinfo {author} {\bibfnamefont {U.~F.}\ \bibnamefont
  {Keyser}},\ }\href {\doibase 10.1002/cphc.201200793} {\bibfield  {journal}
  {\bibinfo  {journal} {Chemphyschem}\ }\textbf {\bibinfo {volume} {14}},\
  \bibinfo {pages} {417} (\bibinfo {year} {2013})}\BibitemShut {NoStop}%
\bibitem [{\citenamefont {Achtyl}\ \emph {et~al.}(2015)\citenamefont {Achtyl},
  \citenamefont {Unocic}, \citenamefont {Xu}, \citenamefont {Cai},
  \citenamefont {Raju}, \citenamefont {Zhang}, \citenamefont {Sacci},
  \citenamefont {Vlassiouk}, \citenamefont {Fulvio}, \citenamefont {Ganesh},
  \citenamefont {Wesolowski}, \citenamefont {Dai}, \citenamefont {van Duin},
  \citenamefont {Neurock},\ and\ \citenamefont {Geiger}}]{Achtyl2015}%
  \BibitemOpen
  \bibfield  {author} {\bibinfo {author} {\bibfnamefont {J.~L.}\ \bibnamefont
  {Achtyl}}, \bibinfo {author} {\bibfnamefont {R.~R.}\ \bibnamefont {Unocic}},
  \bibinfo {author} {\bibfnamefont {L.}~\bibnamefont {Xu}}, \bibinfo {author}
  {\bibfnamefont {Y.}~\bibnamefont {Cai}}, \bibinfo {author} {\bibfnamefont
  {M.}~\bibnamefont {Raju}}, \bibinfo {author} {\bibfnamefont {W.}~\bibnamefont
  {Zhang}}, \bibinfo {author} {\bibfnamefont {R.~L.}\ \bibnamefont {Sacci}},
  \bibinfo {author} {\bibfnamefont {I.~V.}\ \bibnamefont {Vlassiouk}}, \bibinfo
  {author} {\bibfnamefont {P.~F.}\ \bibnamefont {Fulvio}}, \bibinfo {author}
  {\bibfnamefont {P.}~\bibnamefont {Ganesh}}, \bibinfo {author} {\bibfnamefont
  {D.~J.}\ \bibnamefont {Wesolowski}}, \bibinfo {author} {\bibfnamefont
  {S.}~\bibnamefont {Dai}}, \bibinfo {author} {\bibfnamefont {A.~C.~T.}\
  \bibnamefont {van Duin}}, \bibinfo {author} {\bibfnamefont {M.}~\bibnamefont
  {Neurock}}, \ and\ \bibinfo {author} {\bibfnamefont {F.~M.}\ \bibnamefont
  {Geiger}},\ }\href {\doibase 10.1038/ncomms7539} {\bibfield  {journal}
  {\bibinfo  {journal} {Nature Communications}\ }\textbf {\bibinfo {volume}
  {6}},\ \bibinfo {pages} {6539} (\bibinfo {year} {2015})}\BibitemShut
  {NoStop}%
\bibitem [{\citenamefont {Goldman}(1943)}]{Goldman1943}%
  \BibitemOpen
  \bibfield  {author} {\bibinfo {author} {\bibfnamefont {D.~E.}\ \bibnamefont
  {Goldman}},\ }\href@noop {} {\bibfield  {journal} {\bibinfo  {journal}
  {Journal of General Physiology}\ ,\ \bibinfo {pages} {37 }} (\bibinfo {year}
  {1943})}\BibitemShut {NoStop}%
\bibitem [{\citenamefont {Can\c{c}ado}\ \emph {et~al.}(2011)\citenamefont
  {Can\c{c}ado}, \citenamefont {Jorio}, \citenamefont {Ferreira}, \citenamefont
  {Stavale}, \citenamefont {Achete}, \citenamefont {Capaz}, \citenamefont
  {Moutinho}, \citenamefont {Lombardo}, \citenamefont {Kulmala},\ and\
  \citenamefont {Ferrari}}]{Cancado2011}%
  \BibitemOpen
  \bibfield  {author} {\bibinfo {author} {\bibfnamefont {L.~G.}\ \bibnamefont
  {Can\c{c}ado}}, \bibinfo {author} {\bibfnamefont {A.}~\bibnamefont {Jorio}},
  \bibinfo {author} {\bibfnamefont {E.~H.~M.}\ \bibnamefont {Ferreira}},
  \bibinfo {author} {\bibfnamefont {F.}~\bibnamefont {Stavale}}, \bibinfo
  {author} {\bibfnamefont {C.~A.}\ \bibnamefont {Achete}}, \bibinfo {author}
  {\bibfnamefont {R.~B.}\ \bibnamefont {Capaz}}, \bibinfo {author}
  {\bibfnamefont {M.~V.~O.}\ \bibnamefont {Moutinho}}, \bibinfo {author}
  {\bibfnamefont {A.}~\bibnamefont {Lombardo}}, \bibinfo {author}
  {\bibfnamefont {T.~S.}\ \bibnamefont {Kulmala}}, \ and\ \bibinfo {author}
  {\bibfnamefont {A.~C.}\ \bibnamefont {Ferrari}},\ }\href {\doibase
  10.1021/nl201432g} {\bibfield  {journal} {\bibinfo  {journal} {Nano Letters}\
  }\textbf {\bibinfo {volume} {11}},\ \bibinfo {pages} {3190} (\bibinfo {year}
  {2011})}\BibitemShut {NoStop}%
\bibitem [{\citenamefont {Ferrari}\ and\ \citenamefont
  {Basko}(2013)}]{Ferrari2013}%
  \BibitemOpen
  \bibfield  {author} {\bibinfo {author} {\bibfnamefont {A.~C.}\ \bibnamefont
  {Ferrari}}\ and\ \bibinfo {author} {\bibfnamefont {D.~M.}\ \bibnamefont
  {Basko}},\ }\href {\doibase 10.1038/nnano.2013.46} {\bibfield  {journal}
  {\bibinfo  {journal} {Nature Nanotechnology}\ }\textbf {\bibinfo {volume}
  {8}},\ \bibinfo {pages} {235} (\bibinfo {year} {2013})}\BibitemShut {NoStop}%
\bibitem [{\citenamefont {Lucchese}\ \emph {et~al.}(2010)\citenamefont
  {Lucchese}, \citenamefont {Stavale}, \citenamefont {Ferreira}, \citenamefont
  {Vilani}, \citenamefont {Moutinho}, \citenamefont {Capaz}, \citenamefont
  {Achete},\ and\ \citenamefont {Jorio}}]{Lucchese2010}%
  \BibitemOpen
  \bibfield  {author} {\bibinfo {author} {\bibfnamefont {M.}~\bibnamefont
  {Lucchese}}, \bibinfo {author} {\bibfnamefont {F.}~\bibnamefont {Stavale}},
  \bibinfo {author} {\bibfnamefont {E.~M.}\ \bibnamefont {Ferreira}}, \bibinfo
  {author} {\bibfnamefont {C.}~\bibnamefont {Vilani}}, \bibinfo {author}
  {\bibfnamefont {M.}~\bibnamefont {Moutinho}}, \bibinfo {author}
  {\bibfnamefont {R.~B.}\ \bibnamefont {Capaz}}, \bibinfo {author}
  {\bibfnamefont {C.}~\bibnamefont {Achete}}, \ and\ \bibinfo {author}
  {\bibfnamefont {A.}~\bibnamefont {Jorio}},\ }\href {\doibase
  10.1016/j.carbon.2009.12.057} {\bibfield  {journal} {\bibinfo  {journal}
  {Carbon}\ }\textbf {\bibinfo {volume} {48}},\ \bibinfo {pages} {1592}
  (\bibinfo {year} {2010})}\BibitemShut {NoStop}%
\end{thebibliography}%

\end{document}